\documentclass[%
 reprint,
superscriptaddress,
 amsmath,amssymb,
 aps,
 prl,
floatfix,
]{revtex4-2}
\usepackage{cancel}
\usepackage[caption=false]{subfig}
\usepackage{graphicx}% Include figure files
\usepackage{dcolumn}% Align table columns on decimal point
\usepackage{bm}% bold math
\usepackage[hidelinks]{hyperref}% add hypertext capabilities
\usepackage{xcolor}
\usepackage[normalem]{ulem}
\definecolor{ra}{rgb}{0.8, 0.0, 0.0}

\begin{document}

\preprint{APS/123-QED}

\title{Inverse renormalization group of spin glasses}
\author{Dimitrios Bachtis}
\email{dimitrios.bachtis@phys.ens.fr}
\affiliation{Laboratoire de Physique de l'Ecole Normale Sup\'erieure, ENS, Universit\'e PSL,
CNRS, Sorbonne Universit\'e, Universit\'e de Paris, F-75005 Paris, France}

\include{ms.bib}

\date{October 19, 2023}

\begin{abstract}

We propose inverse renormalization group transformations to construct approximate configurations for lattice volumes that have not yet been accessed by  supercomputers or large-scale simulations in the study of spin glasses. Specifically,  starting from lattices of volume $V=8^{3}$ in the case of the three-dimensional Edwards-Anderson model we employ machine learning algorithms to construct rescaled lattices up to $V'=128^{3}$, which we utilize to extract two critical exponents. We conclude by discussing how to incorporate numerical exactness within inverse renormalization group methods of disordered systems, thus opening up the opportunity to explore a sustainable and energy-efficient generation of exact configurations for increasing lattice volumes without the use of dedicated supercomputers.

\end{abstract}

\maketitle

\paragraph*{\label{sec:level1}Introduction.---}

The spin glass~\citep{Mezard1987,RevModPhys.58.801,young1998spin} has posed some of the greatest challenges to the community of statistical mechanics and to computational physicists. Direct Monte Carlo simulations of spin glasses with the Metropolis algorithm fail to thermalize in sufficient computational time as one approaches the spin glass phase. One must then shift to methods which rely on the exchange of configurations between a large number of simulations: these limit the computational resources available to study spin glasses. To exacerbate the above problems, calculations of expectation values for spin glasses require an additional averaging over a large number of realizations of disorder, further limiting the computational power which can be employed to gain insights into the physical behavior of these systems.

Nevertheless, the spin glass is an intrinsically interesting topic. To study spin glasses, and more generally disordered systems, one must investigate the behavior of a system which manifests some form of inconsistency or imperfection in its construction.  Spin glasses have a direct connection with the experimental aspect of science. Disorder, more generally, emerges in all subfields of physics. It is of interest that the above concepts extend beyond physical problems and are relevant for topics such as combinatorial optimization, neural networks, or multi-agent systems~\citep{Mezard1987}. Given the broad impact of spin glasses in the mathematical and physical sciences,  it is of vital importance that we extend the framework of statistical physics in order to accelerate the computational studies of these systems~\citep{PhysRevLett.54.924}.

\begin{figure}[t]
\includegraphics[width=6.5cm]{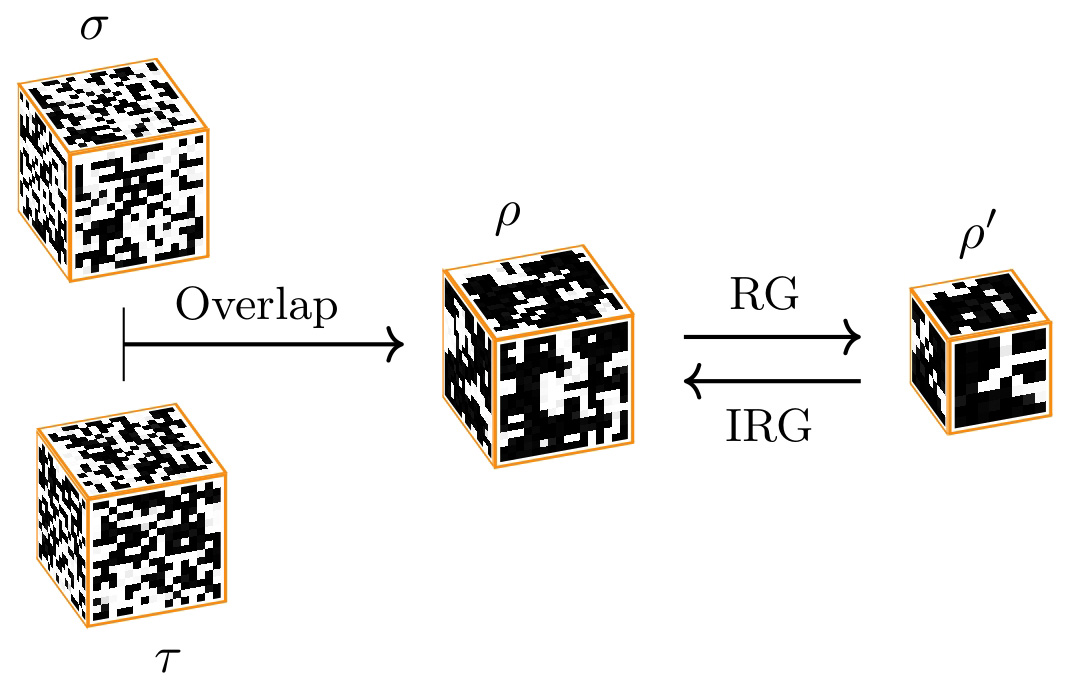}
\caption{\label{fig:irgtrain} The approximate inversion of a standard renormalization group transformation with the use of machine learning. Two replicas $\sigma$, $\tau$ are mapped to an overlap configuration $\rho$ which is renormalized (RG) via the majority rule to obtain $\rho'$. The standard renormalization group is then inverted (IRG) with the use of convolutions to reproduce $\rho$ from $\rho'$. }
\end{figure}

In this Letter we introduce inverse renormalization group transformations~\citep{PhysRevLett.89.275701,PhysRevLett.128.081603,PhysRevB.99.075113,PhysRevLett.121.260601,Shiina2021,waveletrg} to spin glasses. Starting from lattices of volume $V=8^{3}$ in the case of the three-dimensional Edwards-Anderson model we apply a set of inverse transformations to construct lattices of volume $V'=128^{3}$. The method is implemented on an effective Hamiltonian which comprises overlap degrees of freedom~\citep{PhysRevLett.55.2606,PhysRevLett.112.175701,PhysRevB.98.174205,PhysRevB.98.174206}, and approximates the inversion of a standard renormalization group transformation suitable for the study of disordered systems~\citep{Southern_1977,PhysRevB.37.7745,2302.08459} with machine learning algorithms. We additionally explore if the method is capable of producing approximate configurations for lattice volumes that have not yet been accessed by supercomputers~\citep{PhysRevB.88.224416} and large-scale simulations~\citep{PhysRevB.73.224432} in the study of spin glasses.

\begin{figure*}[t]
\includegraphics[width=14cm]{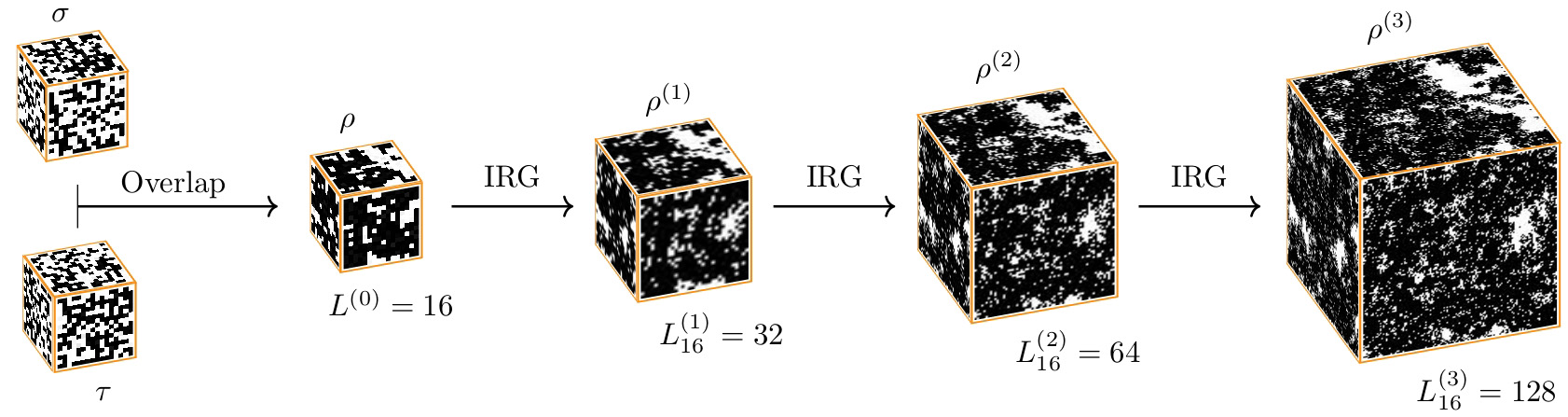}
\caption{\label{fig:irgapply}  Application of the inverse renormalization group. Two replicas $\sigma$, $\tau$ are mapped to an overlap configuration $\rho$. Inverse renormalization group transformations are applied iteratively to construct equilibrated configurations $\rho^{(1)}$, $\rho^{(2)}$ and $\rho^{(3)}$  up to lattice volumes $V=128^{3}$ without requiring additional Monte Carlo simulations on the larger lattices.  }
\end{figure*}

To our knowledge, this Letter documents the first implementation of the inverse renormalization group to study a disordered system. Previous work is limited to applications on exactly-solvable models or systems for which the critical fixed point is known to high numerical accuracy: examples are the Ising model~\citep{PhysRevLett.89.275701,PhysRevB.99.075113,Shiina2021} and the $\phi^{4}$ theory~\citep{PhysRevLett.128.081603}.  In comparison to spin glasses, the aforementioned systems can be easily simulated, in high dimensions and to large lattice volumes. Consequently, we explore if, by generating approximate configurations for lattice volumes that are not accessible by supercomputers and large-scale simulations, this Letter documents the first implemenentation of the inverse renormalization group to obtain a result pertinent to the calculation of critical exponents that is, at the time of writing, inaccessible by other computational means.

To establish the method, we utilize the inversely renormalized configurations to extract two critical exponents based on the Swendsen construction of the renormalization group transformation matrix~\citep{PhysRevLett.42.859} and a two-lattice matching approach~\citep{PhysRevB.37.7745,2302.08459}. We additionally observe the emergence of a critical fixed point via the convergence of the critical exponents using a method originally proposed by Swendsen and Krinsky~\citep{PhysRevLett.43.177}. We conclude by outlining how to incorporate numerical exactness within inverse renormalization group methods of disordered systems, thus opening up the opportunity to explore a sustainable and energy-efficient generation of exact configurations for increasing lattice volumes without the use of dedicated supercomputers.
 
\paragraph*{\label{sec:level2}Transitioning to overlap configurations.---}

We consider the three-dimensional Edwards-Anderson model with helical boundary conditions and two replicas $\sigma$, $\tau$ which comprise spins $s$, $t$. The system is described by the two-replica Hamiltonian:
\begin{equation}\label{eq:origham}
E_{\sigma,\tau} = E_{\sigma}+E_{\tau}=-\sum_{\langle ij \rangle} J_{ij} (s_{i}s_{j}+t_{i}t_{j}),
\end{equation}
where $s,t=\pm 1$, $\langle ij \rangle$ corresponds to nearest-neighbors $i$ and $j$, $J_{ij}$ is a random coupling sampled with equal probability as $J_{ij}=\pm1$, and the set $\lbrace J_{ij} \rbrace$ defines a given realization of disorder for the system.

The Boltzmann probability distribution $p_{\sigma_{i},\tau_{j}}$ of sampling two configurations $\sigma_{i}$, $\tau_{j}$ at inverse temperature $\beta$ is:
\begin{equation}\label{eq:origprob}
p_{\sigma_{i},\tau_{j}}= \frac{\exp[-\beta(E_{\sigma_{i}}+ E_{\tau_{j}})]}{\sum_{\sigma}\sum_{\tau} \exp[-\beta(E_{\sigma}+ E_{\tau})]},
\end{equation}
where $Z^{2}=\sum_{\sigma}\sum_{\tau} \exp[-\beta(E_{\sigma}+ E_{\tau})]$ is the partition function and the sums are over all possible configurations $\sigma$, $\tau$.

The spin glass phase transition of the system is characterized by an overlap order parameter~\citep{PhysRevLett.43.1754,PhysRevLett.50.1946,PhysRevLett.52.1156,Edwards_1975} which is defined over two replicas $\sigma$, $\tau$:
\begin{equation}
q_{\sigma\tau}= \frac{1}{V} \sum_{i} s_{i} t_{i},
\end{equation} 
where $V=L^{3}$ is the volume of the system and $L$ the lattice size in each dimension.

We are now interested in mapping the two-replica Edwards-Anderson model with configurations $\sigma$, $\tau$ to an effective system, with configuration $\rho$, which comprises overlap degrees of freedom $\varrho_{i}=s_{i} t_{i}$. This mapping, introduced by Haake-Lewenstein-Wilkens~\citep{PhysRevLett.55.2606}, defines  an effective Hamiltonian $E^{\textrm{eff}}_{\rho}$, partition function $Z^{\textrm{eff}}_{\rho}$, and Boltzmann probability distribution $P_{\rho_{i}}$ which is averaged over disorder:
\begin{equation}\label{eq:HLW}
P_{\rho_{i}} =   2^{V} \Bigg[  \frac{\exp\big[\beta \sum_{\langle ij \rangle} J_{ij}(1+\varrho_{i}\varrho_{j})\big]}{Z^{2}[\lbrace J_{ij} \rbrace]} \Bigg]_{J_{ij}}.
\end{equation}

We observe that the emergence of the overlap order parameter $q$ in the two-replica Hamiltonian is equivalent to the emergence of a magnetization $m$, summed over the overlap degrees of freedom, in the effective system:
\begin{equation}
m_{\rho}=\frac{1}{V} \sum_{i} \varrho_{i}= \frac{1}{V} \sum_{i} s_{i} t_{i}=q_{\sigma\tau}.
\end{equation}

 Equivalently, the spin glass phase transition of the two-replica Edwards-Anderson model is mapped to a phase transition which resembles ferromagnetic ordering in the effective system.  Our aim is to approximate the inversion of standard renormalization group implementations~\citep{PhysRevB.37.7745,2302.08459} on the overlap degrees of freedom.

\paragraph*{Inverse renormalization group.---}

\begin{figure*}
\includegraphics[width=15.5cm]{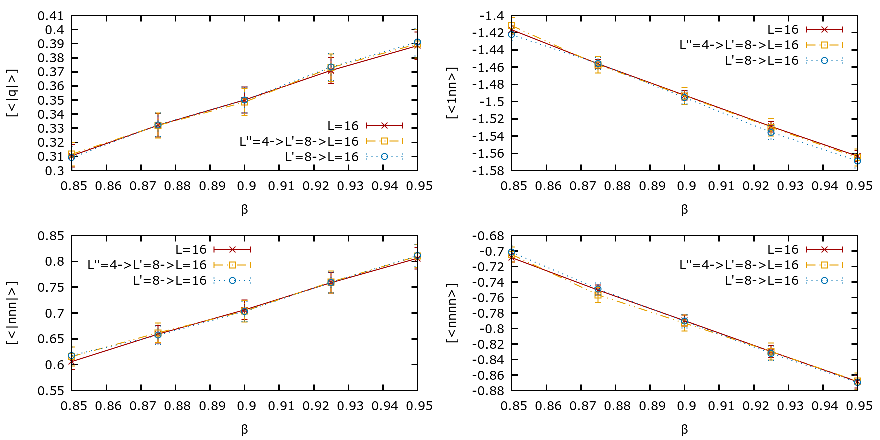}
\caption{\label{fig:all}Observables of the original system of $L=16$ and two inversely renormalized systems $L'=8 \rightarrow L=16$ and $L''=4 \rightarrow L'=8 \rightarrow L=16$ versus the inverse temperature $\beta$. $q$ corresponds to the overlap order parameter, $1nn$ is the interaction of a given spin with the first nearest-neighbor, $nnn$ is a three-spin interaction with the first and second nearest-neighbors, and $nnnn$ is a four-spin interaction with the first, second, and third nearest-neighbors. The machine learning algorithm has not been explicitly trained to reproduce the $nnn$ and $nnnn$ observables.}
\end{figure*}

 We define a standard renormalization group transformation on the effective probability distribution of the overlap configurations as:
\begin{equation}
P'_{\rho'}= \sum_{\rho} T(\rho',\rho) P_{\rho}, 
\end{equation}
where $T(\rho',\rho)$ is a kernel which, in this Letter, corresponds to the majority rule. A standard renormalization group transformation reduces, in terms of lattice units, the lattice size $L$ and the correlation length $\xi$ of the original system by a rescaling factor of $b$, as $L'={L}/{b},$ $ \xi'={\xi}/{b}$.

We approximate the inversion of this standard transformation via the use of machine learning algorithms. Explicitly, we are interested in learning a kernel $\mathcal{T}(\rho,\rho',\lbrace{w}\rbrace)$, where $\lbrace w \rbrace$ is a set of variational parameters, to approximate the reconstruction of configurations of the original system. The implementation of machine learning simply concerns the learning of the most accurate values of $\lbrace w \rbrace$ within the setting of an optimization process to approximate an inversion of the standard transformation. Starting from a renormalized configuration $\rho'$ with lattice size $L'$ and correlation length $\xi'$  we employ the inverse renormalization group to reconstruct an approximation of the original configuration $\rho$ with lattice size $L=bL'$ and correlation length $\xi=b\xi'$. The training process is summarized in Fig.~\ref{fig:irgtrain}. We remark that the application of an inverse renormalization group transformation increases the correlation length of a system by a rescaling factor of $b$ thus driving it closer to the fixed point. 

There exist two advantages in this method. The first is that one can verify, a priori, that a standard renormalization group transformation can be implemented successfully to study the phase transition of the system. As a result, one is able to first optimize the standard renormalization group transformation~\citep{PhysRevLett.52.2321,arxiv.2205.08156}  before one decides to proceed with approximating its inversion. It is already established that the majority rule is a suitable standard transformation for the effective spin glass system~\citep{PhysRevB.37.7745,2302.08459}. The second advantage of the method is that if one considers a machine learning algorithm such as a set of convolutions~\citep{GoodBengCour16,dumoulin2018guide},  which uncover local spatial structure and can be applied irrespective of a given lattice size, then one is able to iteratively increase, in principle for  an arbitrary number of times, the lattice size and the correlation length of the system by a factor of $b$. In this Letter, we consider a rescaling factor of $b=2$. This process, of iteratively increasing the lattice size by a factor of two, is illustrated in Fig.~\ref{fig:irgapply}. 

For the training process we present the  renormalized overlap configurations from $L'=8$ as input to the machine learning algorithm and the original overlap configurations of $L=16$ as the desired output. We consider as a loss function the mean absolute error between multiple observables $S$ of the original and $S'$ of the inversely renormalized system:
\begin{equation}
\mathcal{L}= \sum_{j} \Big|S_{j}-S'_{j}(\lbrace w \rbrace)\Big|.
\end{equation}
%+\sum_{j} \Big|S_{j}(L)-S'_{j}(\lbrace w \rbrace,L'\rightarrow L)\Big|

The loss function provides well-defined standards to approximate the inversion of a renormalization group transformation when assessed from the perspective of the Monte Carlo renormalization group which necessitates the consideration of multiple observables to address emergent systematic uncertainties. The inverse renormalization group approach discussed here considers first the application of a standard renormalization group transformation before proceeding with approximating its inversion. This implies that one can utilize the renormalized configurations from the standard renormalization group implementation to determine the relevant operators and incorporate them as observables in the minimization of the loss function.

We consider as observables the overlap order parameter, and the first, second, and third nearest-neighbor interactions $\rho_{i}\rho_{j}$ as calculated on the overlap degrees of freedom $\rho$. The output of the machine learning algorithm corresponds to the probability that each rescaled degree of freedom $\rho_{i}'$ is assigned the value of $1$. We sample probabilistically the rescaled degrees of freedom, thus verifying that all possible configurations for all possible realizations of disorder have a non-zero probability of appearing. To increase the accuracy of the machine learning implementation we utilize ensemble learning techniques~\citep{Opitz1999} and incorporate boundary conditions on the convolution operations.  Details about the implementation are provided in the Supplemental Material~\footnote{See Supplemental Material at [URL will be inserted by publisher] for details about the machine learning architecture, the renormalization group, and the data analysis}.

In Fig.~\ref{fig:all} we compare observables of original and inversely renormalized systems. Specifically, we invert the renormalization group transformation as $L'=8 \rightarrow L=16$. In addition we consider the case where we apply the majority rule twice on $L=16$ to obtain $L''=4$ and invert the renormalization group twice as $L''=4 \rightarrow L'=8 \rightarrow L=16$. We consider two observables that the machine learning algorithm has been trained on, namely the overlap order parameter and the nearest-neighbor interaction,  and two observables that the machine learning algorithm has not been explicitly trained to reproduce, namely an interaction between three spins and an interaction between four spins. We observe an agreement within statistical uncertainty between the observables of the original system of $L=16$ and those obtained by inverting the transformation. In the Supplemental Material, we demonstrate that the machine learning algorithm can successfully extrapolate beyond the training volumes by reproducing observables for $L''=4\rightarrow L'=8$, and discuss systematic uncertainties pertinent to finite-size effects that manifest in the training process.   

\paragraph*{The critical exponents.---} Via the inverse renormalization group we are able to directly construct approximate configurations of increasing lattice size and correlation length. This implies that it is not necessary to implement Monte Carlo simulations on larger systems to calculate critical exponents. To establish the validity of inverse renormalization group methods in disordered systems we calculate two critical exponents based on the Monte Carlo renormalization group~\citep{PhysRevLett.42.859}, details of which are provided in the Supplemental Material.

To extract the critical exponent $y_{h}$ via the construction of the linearized renormalization group transformation matrix we consider exclusively as an operator the overlap order parameter. We remark that for systems which order ferromagnetically, such as the effective spin glass studied in this Letter, one expects that the consideration of exclusively the order parameter provides an acceptable approximation in the calculation of critical exponents when the considered transformation is the majority rule~\citep{PhysRevB.37.7745}.

% The consideration of further observables in the costruction of the inverse renormalization group transformation matrix, which define odd interactions between the overlap degrees of freedom, would enable an improvement on the obtained results of $y_{h}$ by reducing the pertinent systematic errors. We emphasize that there are well-established standards on how to improve calculations and reduce uncertainties within the Monte Carlo renormalization group~\citep{PhysRevLett.42.859}.
%The results are comparable or in agreement with prior standard renormalization group implementations~\citep{PhysRevB.37.7745,2302.08459}, large-scale simulations~\citep{PhysRevB.73.224432}, and calculations obtained via dedicated supercomputers~\citep{PhysRevB.88.224416}, which simulate lattices up to $V=16^{3}$, $V=24^{3}$, and $V=40^{3}$, respectively

\begin{figure}[t]
\includegraphics[width=8cm]{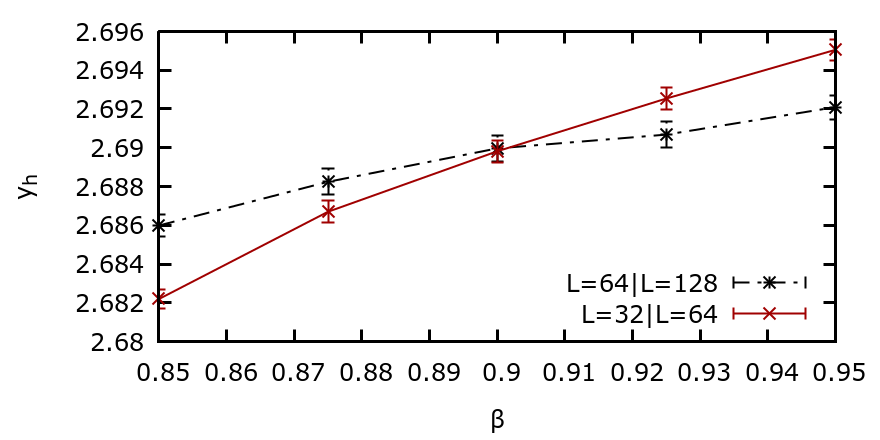}
\caption{\label{fig:exp} Calculation of the critical exponent $y_{h}$ versus the inverse temperature $\beta$. The results are obtained based on the inversely renormalized configurations, starting from $L=16$.}
\end{figure}

\begin{figure}[t]
\includegraphics[width=8cm]{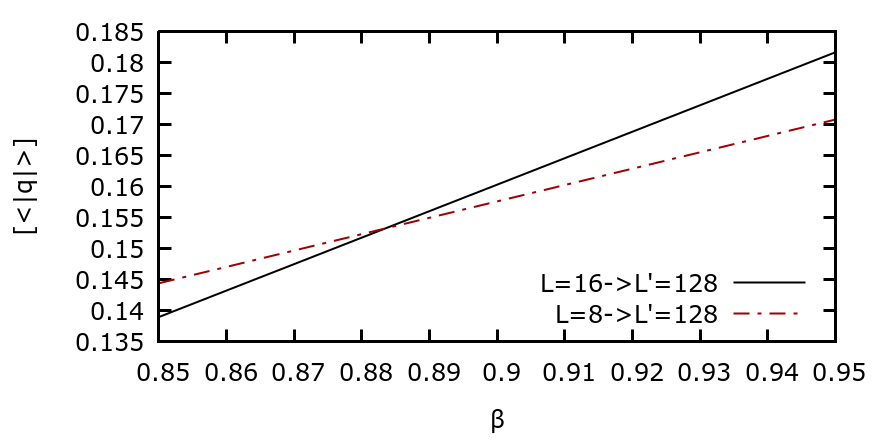}
\caption{\label{fig:expnu} Numerical fits which approximate the linear region in the vicinity of the fixed point using as an observable the overlap order parameter. The fixed point has been obtained by two inversely renormalized systems of identical lattice size $L'=128$, one obtained starting from $L=8$ and another from $L=16$.}
\end{figure}

The calculation of the critical exponent $y_{h}$ is depicted in Fig.~\ref{fig:exp}. The configurations are obtained from an original lattice size of $L=16$ and the iterative application of inverse renormalization group transformations to construct an inversely renormalized lattice of $L'=128$. We recall that, in the context of the inverse renormalization group, the application of an inverse transformation increases the correlation length and drives the system towards the fixed point. Consequently, we expect a convergence of the  critical exponent $y_{h}$ for a wide range of inverse temperatures in the vicinity of the phase transition since the inversely renormalized systems are driven closer to the critical inverse temperature. This observation is verified in Fig.~\ref{fig:exp}.  We remark that the above statements, pertinent to the renormalization group flows, are established on a one-dimensional parameter space under the consideration that the renormalized Hamiltonian is an accurate representation of the original system at a different inverse temperature $\beta'$. Based on the results depicted in Fig.~\ref{fig:exp} for $L=16$ we approximate $y_{h}=2.689(4)$. We then repeat the calculation, starting from $L=8$, and obtain $y_{h}=2.694(2)$. We remark that systematic uncertainties in the calculation of $y_{h}$ can be considered by including additional operators in the construction of the renormalization group matrix.

We now calculate the correlation length exponent $\nu$ using a two lattice matching approach~\citep{PhysRevB.37.7745,2302.08459}. We remark that the calculation of the exponent $\nu$ is numerically intricate since it necessitates the calculation of a gradient in the vicinity of the fixed point. The inverse renormalization group flows therefore affect this calculation. In Fig.~\ref{fig:expnu} we depict numerical fits which approximate the linear region for two inversely renormalized systems of $L'=128$, starting from $L=8\rightarrow L'=128$ and $L=16 \rightarrow L'=128$. By incorporating the statistical uncertainty we calculate the correlation length exponent as $\nu=1.35(5)$ which is in agreement with predictions obtained based on the analytical and numerical treatment of the effective spin glass Hamiltonian studied here~\citep{PhysRevLett.55.2606,PhysRevB.37.7745,PhysRevB.38.9086,2302.08459}. We remark that we do not claim that the calculation of $\nu$ is a conclusive result for the exponent of the original spin glass Hamiltonian, only that it is consistent with prior predictions using configurations of the effective spin glass. The calculation can be improved by considering the fixed points which emerge from other observables. We remark that we have not assumed a convergence to the renormalized trajectory for the results obtained in this Letter. 

\paragraph*{Conclusions.---} 
 
We have introduced inverse renormalization group methods to spin glasses and to disordered systems. Specifically, starting from lattices of volume $V=8^{3}$ in the case of the three-dimensional Edwards-Anderson model we employed a set of inverse transformations to construct approximate configurations for lattices of volume $V=128^{3}$. Inverse renormalization group methods overcome the critical slowing down effect in the calculation of critical exponents~\citep{PhysRevLett.89.275701,PhysRevLett.128.081603,waveletrg}. Since the method  enables the generation  of approximate configurations for lattice volumes that have not yet been accessed by supercomputers or large-scale simulations it can be implemented to predict, under a significant reduction of finite-size effects, the critical behavior of a system before it becomes known via first-principles calculations.

 There exist three impediments to be overcome in order to recast the inverse renormalization group of disordered systems as a numerically exact method and produce exact configurations of the inversely renormalized system. These are the numerical determination of the Hamiltonian for the effective system of Eq.~\ref{eq:HLW}, the conception of an inverse transformation for the overlap degrees of freedom, and the proposal of an inverse renormalization for the effective realization of disorder. The numerical determination of the Hamiltonian for the effective system is already solved by Wang and Swendsen~\citep{PhysRevB.38.9086}. The current Letter proposes a solution to the conception of an inverse transformation for the overlap degrees of freedom. Consequently, the remaining requirement in order to establish numerical exactness is to devise an inverse renormalization for the effective couplings. Once this is achieved one can introduce the configurations as proposed moves within a Monte Carlo setting: we remark that such studies are heavily dependent on the choice of the standard and inverse transformation which give rise to a certain space of renormalized Hamiltonians. 

By incorporating numerical exactness, the inverse renormalization group has the potential to achieve a sustainable and energy-efficient generation of exact configurations for lattice sizes that are inaccessible by supercomputers or large-scale simulations, thus providing substantial computational benefits in comparison to conventional Monte Carlo simulations within one of the most numerically challenging research fields of physics, namely the theory of disordered systems.

The author thanks Giulio Biroli for interesting discussions and acknowledges support from the CFM-ENS Data Science Chair and PRAIRIE (the PaRis Artificial Intelligence Research InstitutE).
%\clearpage

\bibliography{ms}

\providecommand{\noopsort}[1]{}\providecommand{\singleletter}[1]{#1}%
\begin{thebibliography}{32}%
\makeatletter
\providecommand \@ifxundefined [1]{%
 \@ifx{#1\undefined}
}%
\providecommand \@ifnum [1]{%
 \ifnum #1\expandafter \@firstoftwo
 \else \expandafter \@secondoftwo
 \fi
}%
\providecommand \@ifx [1]{%
 \ifx #1\expandafter \@firstoftwo
 \else \expandafter \@secondoftwo
 \fi
}%
\providecommand \natexlab [1]{#1}%
\providecommand \enquote  [1]{``#1''}%
\providecommand \bibnamefont  [1]{#1}%
\providecommand \bibfnamefont [1]{#1}%
\providecommand \citenamefont [1]{#1}%
\providecommand \href@noop [0]{\@secondoftwo}%
\providecommand \href [0]{\begingroup \@sanitize@url \@href}%
\providecommand \@href[1]{\@@startlink{#1}\@@href}%
\providecommand \@@href[1]{\endgroup#1\@@endlink}%
\providecommand \@sanitize@url [0]{\catcode `\\12\catcode `\$12\catcode
  `\&12\catcode `\#12\catcode `\^12\catcode `\_12\catcode `\%12\relax}%
\providecommand \@@startlink[1]{}%
\providecommand \@@endlink[0]{}%
\providecommand \url  [0]{\begingroup\@sanitize@url \@url }%
\providecommand \@url [1]{\endgroup\@href {#1}{\urlprefix }}%
\providecommand \urlprefix  [0]{URL }%
\providecommand \Eprint [0]{\href }%
\providecommand \doibase [0]{https://doi.org/}%
\providecommand \selectlanguage [0]{\@gobble}%
\providecommand \bibinfo  [0]{\@secondoftwo}%
\providecommand \bibfield  [0]{\@secondoftwo}%
\providecommand \translation [1]{[#1]}%
\providecommand \BibitemOpen [0]{}%
\providecommand \bibitemStop [0]{}%
\providecommand \bibitemNoStop [0]{.\EOS\space}%
\providecommand \EOS [0]{\spacefactor3000\relax}%
\providecommand \BibitemShut  [1]{\csname bibitem#1\endcsname}%
\let\auto@bib@innerbib\@empty
%</preamble>
\bibitem [{\citenamefont {M{\'e}zard}\ \emph {et~al.}(1987)\citenamefont
  {M{\'e}zard}, \citenamefont {Parisi},\ and\ \citenamefont
  {Virasoro}}]{Mezard1987}%
  \BibitemOpen
  \bibfield  {author} {\bibinfo {author} {\bibfnamefont {M.}~\bibnamefont
  {M{\'e}zard}}, \bibinfo {author} {\bibfnamefont {G.}~\bibnamefont {Parisi}},\
  and\ \bibinfo {author} {\bibfnamefont {M.~A.}\ \bibnamefont {Virasoro}},\
  }\href@noop {} {\emph {\bibinfo {title} {Spin Glass Theory and Beyond: An
  Introduction to the Replica Method and Its Applications}}}\ (\bibinfo
  {publisher} {World Scientific},\ \bibinfo {address} {Singapore},\ \bibinfo
  {year} {1987})\BibitemShut {NoStop}%
\bibitem [{\citenamefont {Binder}\ and\ \citenamefont
  {Young}(1986)}]{RevModPhys.58.801}%
  \BibitemOpen
  \bibfield  {author} {\bibinfo {author} {\bibfnamefont {K.}~\bibnamefont
  {Binder}}\ and\ \bibinfo {author} {\bibfnamefont {A.~P.}\ \bibnamefont
  {Young}},\ }\bibfield  {title} {\bibinfo {title} {Spin glasses: Experimental
  facts, theoretical concepts, and open questions},\ }\href
  {https://doi.org/10.1103/RevModPhys.58.801} {\bibfield  {journal} {\bibinfo
  {journal} {Rev. Mod. Phys.}\ }\textbf {\bibinfo {volume} {58}},\ \bibinfo
  {pages} {801} (\bibinfo {year} {1986})}\BibitemShut {NoStop}%
\bibitem [{\citenamefont {Young}(1998)}]{young1998spin}%
  \BibitemOpen
  \bibfield  {author} {\bibinfo {author} {\bibfnamefont {A.}~\bibnamefont
  {Young}},\ }\href@noop {} {\emph {\bibinfo {title} {Spin Glasses and Random
  Fields}}},\ Directions in condensed matter physics\ (\bibinfo  {publisher}
  {World Scientific},\ \bibinfo {year} {1998})\BibitemShut {NoStop}%
\bibitem [{\citenamefont {Bhatt}\ and\ \citenamefont
  {Young}(1985)}]{PhysRevLett.54.924}%
  \BibitemOpen
  \bibfield  {author} {\bibinfo {author} {\bibfnamefont {R.~N.}\ \bibnamefont
  {Bhatt}}\ and\ \bibinfo {author} {\bibfnamefont {A.~P.}\ \bibnamefont
  {Young}},\ }\bibfield  {title} {\bibinfo {title} {Search for a transition in
  the three-dimensional j ising spin-glass},\ }\href
  {https://doi.org/10.1103/PhysRevLett.54.924} {\bibfield  {journal} {\bibinfo
  {journal} {Phys. Rev. Lett.}\ }\textbf {\bibinfo {volume} {54}},\ \bibinfo
  {pages} {924} (\bibinfo {year} {1985})}\BibitemShut {NoStop}%
\bibitem [{\citenamefont {Ron}\ \emph {et~al.}(2002)\citenamefont {Ron},
  \citenamefont {Swendsen},\ and\ \citenamefont
  {Brandt}}]{PhysRevLett.89.275701}%
  \BibitemOpen
  \bibfield  {author} {\bibinfo {author} {\bibfnamefont {D.}~\bibnamefont
  {Ron}}, \bibinfo {author} {\bibfnamefont {R.~H.}\ \bibnamefont {Swendsen}},\
  and\ \bibinfo {author} {\bibfnamefont {A.}~\bibnamefont {Brandt}},\
  }\bibfield  {title} {\bibinfo {title} {Inverse monte carlo renormalization
  group transformations for critical phenomena},\ }\href
  {https://doi.org/10.1103/PhysRevLett.89.275701} {\bibfield  {journal}
  {\bibinfo  {journal} {Phys. Rev. Lett.}\ }\textbf {\bibinfo {volume} {89}},\
  \bibinfo {pages} {275701} (\bibinfo {year} {2002})}\BibitemShut {NoStop}%
\bibitem [{\citenamefont {Bachtis}\ \emph {et~al.}(2022)\citenamefont
  {Bachtis}, \citenamefont {Aarts}, \citenamefont {Di~Renzo},\ and\
  \citenamefont {Lucini}}]{PhysRevLett.128.081603}%
  \BibitemOpen
  \bibfield  {author} {\bibinfo {author} {\bibfnamefont {D.}~\bibnamefont
  {Bachtis}}, \bibinfo {author} {\bibfnamefont {G.}~\bibnamefont {Aarts}},
  \bibinfo {author} {\bibfnamefont {F.}~\bibnamefont {Di~Renzo}},\ and\
  \bibinfo {author} {\bibfnamefont {B.}~\bibnamefont {Lucini}},\ }\bibfield
  {title} {\bibinfo {title} {Inverse renormalization group in quantum field
  theory},\ }\href {https://doi.org/10.1103/PhysRevLett.128.081603} {\bibfield
  {journal} {\bibinfo  {journal} {Phys. Rev. Lett.}\ }\textbf {\bibinfo
  {volume} {128}},\ \bibinfo {pages} {081603} (\bibinfo {year}
  {2022})}\BibitemShut {NoStop}%
\bibitem [{\citenamefont {Efthymiou}\ \emph {et~al.}(2019)\citenamefont
  {Efthymiou}, \citenamefont {Beach},\ and\ \citenamefont
  {Melko}}]{PhysRevB.99.075113}%
  \BibitemOpen
  \bibfield  {author} {\bibinfo {author} {\bibfnamefont {S.}~\bibnamefont
  {Efthymiou}}, \bibinfo {author} {\bibfnamefont {M.~J.~S.}\ \bibnamefont
  {Beach}},\ and\ \bibinfo {author} {\bibfnamefont {R.~G.}\ \bibnamefont
  {Melko}},\ }\bibfield  {title} {\bibinfo {title} {Super-resolving the ising
  model with convolutional neural networks},\ }\href
  {https://doi.org/10.1103/PhysRevB.99.075113} {\bibfield  {journal} {\bibinfo
  {journal} {Phys. Rev. B}\ }\textbf {\bibinfo {volume} {99}},\ \bibinfo
  {pages} {075113} (\bibinfo {year} {2019})}\BibitemShut {NoStop}%
\bibitem [{\citenamefont {Li}\ and\ \citenamefont
  {Wang}(2018)}]{PhysRevLett.121.260601}%
  \BibitemOpen
  \bibfield  {author} {\bibinfo {author} {\bibfnamefont {S.-H.}\ \bibnamefont
  {Li}}\ and\ \bibinfo {author} {\bibfnamefont {L.}~\bibnamefont {Wang}},\
  }\bibfield  {title} {\bibinfo {title} {Neural network renormalization
  group},\ }\href {https://doi.org/10.1103/PhysRevLett.121.260601} {\bibfield
  {journal} {\bibinfo  {journal} {Phys. Rev. Lett.}\ }\textbf {\bibinfo
  {volume} {121}},\ \bibinfo {pages} {260601} (\bibinfo {year}
  {2018})}\BibitemShut {NoStop}%
\bibitem [{\citenamefont {Shiina}\ \emph {et~al.}(2021)\citenamefont {Shiina},
  \citenamefont {Mori}, \citenamefont {Tomita}, \citenamefont {Lee},\ and\
  \citenamefont {Okabe}}]{Shiina2021}%
  \BibitemOpen
  \bibfield  {author} {\bibinfo {author} {\bibfnamefont {K.}~\bibnamefont
  {Shiina}}, \bibinfo {author} {\bibfnamefont {H.}~\bibnamefont {Mori}},
  \bibinfo {author} {\bibfnamefont {Y.}~\bibnamefont {Tomita}}, \bibinfo
  {author} {\bibfnamefont {H.~K.}\ \bibnamefont {Lee}},\ and\ \bibinfo {author}
  {\bibfnamefont {Y.}~\bibnamefont {Okabe}},\ }\bibfield  {title} {\bibinfo
  {title} {Inverse renormalization group based on image super-resolution using
  deep convolutional networks},\ }\href
  {https://doi.org/10.1038/s41598-021-88605-w} {\bibfield  {journal} {\bibinfo
  {journal} {Scientific Reports}\ }\textbf {\bibinfo {volume} {11}},\ \bibinfo
  {pages} {9617} (\bibinfo {year} {2021})}\BibitemShut {NoStop}%
\bibitem [{\citenamefont {Marchand}\ \emph {et~al.}(2022)\citenamefont
  {Marchand}, \citenamefont {Ozawa}, \citenamefont {Biroli},\ and\
  \citenamefont {Mallat}}]{waveletrg}%
  \BibitemOpen
  \bibfield  {author} {\bibinfo {author} {\bibfnamefont {T.}~\bibnamefont
  {Marchand}}, \bibinfo {author} {\bibfnamefont {M.}~\bibnamefont {Ozawa}},
  \bibinfo {author} {\bibfnamefont {G.}~\bibnamefont {Biroli}},\ and\ \bibinfo
  {author} {\bibfnamefont {S.}~\bibnamefont {Mallat}},\ }\href
  {https://doi.org/10.48550/ARXIV.2207.04941} {\bibinfo {title} {Wavelet
  conditional renormalization group}} (\bibinfo {year} {2022})\BibitemShut
  {NoStop}%
\bibitem [{\citenamefont {Haake}\ \emph {et~al.}(1985)\citenamefont {Haake},
  \citenamefont {Lewenstein},\ and\ \citenamefont
  {Wilkens}}]{PhysRevLett.55.2606}%
  \BibitemOpen
  \bibfield  {author} {\bibinfo {author} {\bibfnamefont {F.}~\bibnamefont
  {Haake}}, \bibinfo {author} {\bibfnamefont {M.}~\bibnamefont {Lewenstein}},\
  and\ \bibinfo {author} {\bibfnamefont {M.}~\bibnamefont {Wilkens}},\
  }\bibfield  {title} {\bibinfo {title} {Relation of random and competing
  nonrandom couplings for spin-glasses},\ }\href
  {https://doi.org/10.1103/PhysRevLett.55.2606} {\bibfield  {journal} {\bibinfo
   {journal} {Phys. Rev. Lett.}\ }\textbf {\bibinfo {volume} {55}},\ \bibinfo
  {pages} {2606} (\bibinfo {year} {1985})}\BibitemShut {NoStop}%
\bibitem [{\citenamefont {Biroli}\ \emph {et~al.}(2014)\citenamefont {Biroli},
  \citenamefont {Cammarota}, \citenamefont {Tarjus},\ and\ \citenamefont
  {Tarzia}}]{PhysRevLett.112.175701}%
  \BibitemOpen
  \bibfield  {author} {\bibinfo {author} {\bibfnamefont {G.}~\bibnamefont
  {Biroli}}, \bibinfo {author} {\bibfnamefont {C.}~\bibnamefont {Cammarota}},
  \bibinfo {author} {\bibfnamefont {G.}~\bibnamefont {Tarjus}},\ and\ \bibinfo
  {author} {\bibfnamefont {M.}~\bibnamefont {Tarzia}},\ }\bibfield  {title}
  {\bibinfo {title} {Random-field-like criticality in glass-forming liquids},\
  }\href {https://doi.org/10.1103/PhysRevLett.112.175701} {\bibfield  {journal}
  {\bibinfo  {journal} {Phys. Rev. Lett.}\ }\textbf {\bibinfo {volume} {112}},\
  \bibinfo {pages} {175701} (\bibinfo {year} {2014})}\BibitemShut {NoStop}%
\bibitem [{\citenamefont {Biroli}\ \emph
  {et~al.}(2018{\natexlab{a}})\citenamefont {Biroli}, \citenamefont
  {Cammarota}, \citenamefont {Tarjus},\ and\ \citenamefont
  {Tarzia}}]{PhysRevB.98.174205}%
  \BibitemOpen
  \bibfield  {author} {\bibinfo {author} {\bibfnamefont {G.}~\bibnamefont
  {Biroli}}, \bibinfo {author} {\bibfnamefont {C.}~\bibnamefont {Cammarota}},
  \bibinfo {author} {\bibfnamefont {G.}~\bibnamefont {Tarjus}},\ and\ \bibinfo
  {author} {\bibfnamefont {M.}~\bibnamefont {Tarzia}},\ }\bibfield  {title}
  {\bibinfo {title} {Random-field ising-like effective theory of the glass
  transition. i. mean-field models},\ }\href
  {https://doi.org/10.1103/PhysRevB.98.174205} {\bibfield  {journal} {\bibinfo
  {journal} {Phys. Rev. B}\ }\textbf {\bibinfo {volume} {98}},\ \bibinfo
  {pages} {174205} (\bibinfo {year} {2018}{\natexlab{a}})}\BibitemShut
  {NoStop}%
\bibitem [{\citenamefont {Biroli}\ \emph
  {et~al.}(2018{\natexlab{b}})\citenamefont {Biroli}, \citenamefont
  {Cammarota}, \citenamefont {Tarjus},\ and\ \citenamefont
  {Tarzia}}]{PhysRevB.98.174206}%
  \BibitemOpen
  \bibfield  {author} {\bibinfo {author} {\bibfnamefont {G.}~\bibnamefont
  {Biroli}}, \bibinfo {author} {\bibfnamefont {C.}~\bibnamefont {Cammarota}},
  \bibinfo {author} {\bibfnamefont {G.}~\bibnamefont {Tarjus}},\ and\ \bibinfo
  {author} {\bibfnamefont {M.}~\bibnamefont {Tarzia}},\ }\bibfield  {title}
  {\bibinfo {title} {Random field ising-like effective theory of the glass
  transition. ii. finite-dimensional models},\ }\href
  {https://doi.org/10.1103/PhysRevB.98.174206} {\bibfield  {journal} {\bibinfo
  {journal} {Phys. Rev. B}\ }\textbf {\bibinfo {volume} {98}},\ \bibinfo
  {pages} {174206} (\bibinfo {year} {2018}{\natexlab{b}})}\BibitemShut
  {NoStop}%
\bibitem [{\citenamefont {Southern}\ and\ \citenamefont
  {Young}(1977)}]{Southern_1977}%
  \BibitemOpen
  \bibfield  {author} {\bibinfo {author} {\bibfnamefont {B.~W.}\ \bibnamefont
  {Southern}}\ and\ \bibinfo {author} {\bibfnamefont {A.~P.}\ \bibnamefont
  {Young}},\ }\bibfield  {title} {\bibinfo {title} {Real space rescaling study
  of spin glass behaviour in three dimensions},\ }\href
  {https://doi.org/10.1088/0022-3719/10/12/023} {\bibfield  {journal} {\bibinfo
   {journal} {Journal of Physics C: Solid State Physics}\ }\textbf {\bibinfo
  {volume} {10}},\ \bibinfo {pages} {2179} (\bibinfo {year}
  {1977})}\BibitemShut {NoStop}%
\bibitem [{\citenamefont {Wang}\ and\ \citenamefont
  {Swendsen}(1988{\natexlab{a}})}]{PhysRevB.37.7745}%
  \BibitemOpen
  \bibfield  {author} {\bibinfo {author} {\bibfnamefont {J.-S.}\ \bibnamefont
  {Wang}}\ and\ \bibinfo {author} {\bibfnamefont {R.~H.}\ \bibnamefont
  {Swendsen}},\ }\bibfield  {title} {\bibinfo {title} {Monte carlo
  renormalization-group study of ising spin glasses},\ }\href
  {https://doi.org/10.1103/PhysRevB.37.7745} {\bibfield  {journal} {\bibinfo
  {journal} {Phys. Rev. B}\ }\textbf {\bibinfo {volume} {37}},\ \bibinfo
  {pages} {7745} (\bibinfo {year} {1988}{\natexlab{a}})}\BibitemShut {NoStop}%
\bibitem [{\citenamefont {Bachtis}(2024{\natexlab{a}})}]{2302.08459}%
  \BibitemOpen
  \bibfield  {author} {\bibinfo {author} {\bibfnamefont {D.}~\bibnamefont
  {Bachtis}},\ }\bibfield  {title} {\bibinfo {title} {Overlap renormalization
  group transformations for disordered systems},\ }\href
  {http://iopscience.iop.org/article/10.1088/1751-8121/ad4c2e} {\bibfield
  {journal} {\bibinfo  {journal} {Journal of Physics A: Mathematical and
  Theoretical}\ } (\bibinfo {year} {2024}{\natexlab{a}})}\BibitemShut {NoStop}%
\bibitem [{\citenamefont {Baity-Jesi}\ \emph {et~al.}(2013)\citenamefont
  {Baity-Jesi}, \citenamefont {Ba\~nos}, \citenamefont {Cruz}, \citenamefont
  {Fernandez}, \citenamefont {Gil-Narvion}, \citenamefont {Gordillo-Guerrero},
  \citenamefont {I\~niguez}, \citenamefont {Maiorano}, \citenamefont
  {Mantovani}, \citenamefont {Marinari}, \citenamefont {Martin-Mayor},
  \citenamefont {Monforte-Garcia}, \citenamefont {Sudupe}, \citenamefont
  {Navarro}, \citenamefont {Parisi}, \citenamefont {Perez-Gaviro},
  \citenamefont {Pivanti}, \citenamefont {Ricci-Tersenghi}, \citenamefont
  {Ruiz-Lorenzo}, \citenamefont {Schifano}, \citenamefont {Seoane},
  \citenamefont {Tarancon}, \citenamefont {Tripiccione},\ and\ \citenamefont
  {Yllanes}}]{PhysRevB.88.224416}%
  \BibitemOpen
  \bibfield  {author} {\bibinfo {author} {\bibfnamefont {M.}~\bibnamefont
  {Baity-Jesi}}, \bibinfo {author} {\bibfnamefont {R.~A.}\ \bibnamefont
  {Ba\~nos}}, \bibinfo {author} {\bibfnamefont {A.}~\bibnamefont {Cruz}},
  \bibinfo {author} {\bibfnamefont {L.~A.}\ \bibnamefont {Fernandez}}, \bibinfo
  {author} {\bibfnamefont {J.~M.}\ \bibnamefont {Gil-Narvion}}, \bibinfo
  {author} {\bibfnamefont {A.}~\bibnamefont {Gordillo-Guerrero}}, \bibinfo
  {author} {\bibfnamefont {D.}~\bibnamefont {I\~niguez}}, \bibinfo {author}
  {\bibfnamefont {A.}~\bibnamefont {Maiorano}}, \bibinfo {author}
  {\bibfnamefont {F.}~\bibnamefont {Mantovani}}, \bibinfo {author}
  {\bibfnamefont {E.}~\bibnamefont {Marinari}}, \bibinfo {author}
  {\bibfnamefont {V.}~\bibnamefont {Martin-Mayor}}, \bibinfo {author}
  {\bibfnamefont {J.}~\bibnamefont {Monforte-Garcia}}, \bibinfo {author}
  {\bibfnamefont {A.~M.~n.}\ \bibnamefont {Sudupe}}, \bibinfo {author}
  {\bibfnamefont {D.}~\bibnamefont {Navarro}}, \bibinfo {author} {\bibfnamefont
  {G.}~\bibnamefont {Parisi}}, \bibinfo {author} {\bibfnamefont
  {S.}~\bibnamefont {Perez-Gaviro}}, \bibinfo {author} {\bibfnamefont
  {M.}~\bibnamefont {Pivanti}}, \bibinfo {author} {\bibfnamefont
  {F.}~\bibnamefont {Ricci-Tersenghi}}, \bibinfo {author} {\bibfnamefont
  {J.~J.}\ \bibnamefont {Ruiz-Lorenzo}}, \bibinfo {author} {\bibfnamefont
  {S.~F.}\ \bibnamefont {Schifano}}, \bibinfo {author} {\bibfnamefont
  {B.}~\bibnamefont {Seoane}}, \bibinfo {author} {\bibfnamefont
  {A.}~\bibnamefont {Tarancon}}, \bibinfo {author} {\bibfnamefont
  {R.}~\bibnamefont {Tripiccione}},\ and\ \bibinfo {author} {\bibfnamefont
  {D.}~\bibnamefont {Yllanes}} (\bibinfo {collaboration} {Janus
  Collaboration}),\ }\bibfield  {title} {\bibinfo {title} {Critical parameters
  of the three-dimensional ising spin glass},\ }\href
  {https://doi.org/10.1103/PhysRevB.88.224416} {\bibfield  {journal} {\bibinfo
  {journal} {Phys. Rev. B}\ }\textbf {\bibinfo {volume} {88}},\ \bibinfo
  {pages} {224416} (\bibinfo {year} {2013})}\BibitemShut {NoStop}%
\bibitem [{\citenamefont {Katzgraber}\ \emph {et~al.}(2006)\citenamefont
  {Katzgraber}, \citenamefont {K\"orner},\ and\ \citenamefont
  {Young}}]{PhysRevB.73.224432}%
  \BibitemOpen
  \bibfield  {author} {\bibinfo {author} {\bibfnamefont {H.~G.}\ \bibnamefont
  {Katzgraber}}, \bibinfo {author} {\bibfnamefont {M.}~\bibnamefont
  {K\"orner}},\ and\ \bibinfo {author} {\bibfnamefont {A.~P.}\ \bibnamefont
  {Young}},\ }\bibfield  {title} {\bibinfo {title} {Universality in
  three-dimensional ising spin glasses: A monte carlo study},\ }\href
  {https://doi.org/10.1103/PhysRevB.73.224432} {\bibfield  {journal} {\bibinfo
  {journal} {Phys. Rev. B}\ }\textbf {\bibinfo {volume} {73}},\ \bibinfo
  {pages} {224432} (\bibinfo {year} {2006})}\BibitemShut {NoStop}%
\bibitem [{\citenamefont {Swendsen}(1979)}]{PhysRevLett.42.859}%
  \BibitemOpen
  \bibfield  {author} {\bibinfo {author} {\bibfnamefont {R.~H.}\ \bibnamefont
  {Swendsen}},\ }\bibfield  {title} {\bibinfo {title} {Monte carlo
  renormalization group},\ }\href {https://doi.org/10.1103/PhysRevLett.42.859}
  {\bibfield  {journal} {\bibinfo  {journal} {Phys. Rev. Lett.}\ }\textbf
  {\bibinfo {volume} {42}},\ \bibinfo {pages} {859} (\bibinfo {year}
  {1979})}\BibitemShut {NoStop}%
\bibitem [{\citenamefont {Swendsen}\ and\ \citenamefont
  {Krinsky}(1979)}]{PhysRevLett.43.177}%
  \BibitemOpen
  \bibfield  {author} {\bibinfo {author} {\bibfnamefont {R.~H.}\ \bibnamefont
  {Swendsen}}\ and\ \bibinfo {author} {\bibfnamefont {S.}~\bibnamefont
  {Krinsky}},\ }\bibfield  {title} {\bibinfo {title} {Monte carlo
  renormalization group and ising models with n>~2},\ }\href
  {https://doi.org/10.1103/PhysRevLett.43.177} {\bibfield  {journal} {\bibinfo
  {journal} {Phys. Rev. Lett.}\ }\textbf {\bibinfo {volume} {43}},\ \bibinfo
  {pages} {177} (\bibinfo {year} {1979})}\BibitemShut {NoStop}%
\bibitem [{\citenamefont {Parisi}(1979)}]{PhysRevLett.43.1754}%
  \BibitemOpen
  \bibfield  {author} {\bibinfo {author} {\bibfnamefont {G.}~\bibnamefont
  {Parisi}},\ }\bibfield  {title} {\bibinfo {title} {Infinite number of order
  parameters for spin-glasses},\ }\href
  {https://doi.org/10.1103/PhysRevLett.43.1754} {\bibfield  {journal} {\bibinfo
   {journal} {Phys. Rev. Lett.}\ }\textbf {\bibinfo {volume} {43}},\ \bibinfo
  {pages} {1754} (\bibinfo {year} {1979})}\BibitemShut {NoStop}%
\bibitem [{\citenamefont {Parisi}(1983)}]{PhysRevLett.50.1946}%
  \BibitemOpen
  \bibfield  {author} {\bibinfo {author} {\bibfnamefont {G.}~\bibnamefont
  {Parisi}},\ }\bibfield  {title} {\bibinfo {title} {Order parameter for
  spin-glasses},\ }\href {https://doi.org/10.1103/PhysRevLett.50.1946}
  {\bibfield  {journal} {\bibinfo  {journal} {Phys. Rev. Lett.}\ }\textbf
  {\bibinfo {volume} {50}},\ \bibinfo {pages} {1946} (\bibinfo {year}
  {1983})}\BibitemShut {NoStop}%
\bibitem [{\citenamefont {M\'ezard}\ \emph {et~al.}(1984)\citenamefont
  {M\'ezard}, \citenamefont {Parisi}, \citenamefont {Sourlas}, \citenamefont
  {Toulouse},\ and\ \citenamefont {Virasoro}}]{PhysRevLett.52.1156}%
  \BibitemOpen
  \bibfield  {author} {\bibinfo {author} {\bibfnamefont {M.}~\bibnamefont
  {M\'ezard}}, \bibinfo {author} {\bibfnamefont {G.}~\bibnamefont {Parisi}},
  \bibinfo {author} {\bibfnamefont {N.}~\bibnamefont {Sourlas}}, \bibinfo
  {author} {\bibfnamefont {G.}~\bibnamefont {Toulouse}},\ and\ \bibinfo
  {author} {\bibfnamefont {M.}~\bibnamefont {Virasoro}},\ }\bibfield  {title}
  {\bibinfo {title} {Nature of the spin-glass phase},\ }\href
  {https://doi.org/10.1103/PhysRevLett.52.1156} {\bibfield  {journal} {\bibinfo
   {journal} {Phys. Rev. Lett.}\ }\textbf {\bibinfo {volume} {52}},\ \bibinfo
  {pages} {1156} (\bibinfo {year} {1984})}\BibitemShut {NoStop}%
\bibitem [{\citenamefont {Edwards}\ and\ \citenamefont
  {Anderson}(1975)}]{Edwards_1975}%
  \BibitemOpen
  \bibfield  {author} {\bibinfo {author} {\bibfnamefont {S.~F.}\ \bibnamefont
  {Edwards}}\ and\ \bibinfo {author} {\bibfnamefont {P.~W.}\ \bibnamefont
  {Anderson}},\ }\bibfield  {title} {\bibinfo {title} {Theory of spin
  glasses},\ }\href {https://doi.org/10.1088/0305-4608/5/5/017} {\bibfield
  {journal} {\bibinfo  {journal} {Journal of Physics F: Metal Physics}\
  }\textbf {\bibinfo {volume} {5}},\ \bibinfo {pages} {965} (\bibinfo {year}
  {1975})}\BibitemShut {NoStop}%
\bibitem [{\citenamefont {Swendsen}(1984)}]{PhysRevLett.52.2321}%
  \BibitemOpen
  \bibfield  {author} {\bibinfo {author} {\bibfnamefont {R.~H.}\ \bibnamefont
  {Swendsen}},\ }\bibfield  {title} {\bibinfo {title} {Optimization of
  real-space renormalization-group transformations},\ }\href
  {https://doi.org/10.1103/PhysRevLett.52.2321} {\bibfield  {journal} {\bibinfo
   {journal} {Phys. Rev. Lett.}\ }\textbf {\bibinfo {volume} {52}},\ \bibinfo
  {pages} {2321} (\bibinfo {year} {1984})}\BibitemShut {NoStop}%
\bibitem [{\citenamefont {Bachtis}(2024{\natexlab{b}})}]{arxiv.2205.08156}%
  \BibitemOpen
  \bibfield  {author} {\bibinfo {author} {\bibfnamefont {D.}~\bibnamefont
  {Bachtis}},\ }\bibfield  {title} {\bibinfo {title} {Reducing finite-size
  effects with reweighted renormalization group transformations},\ }\href
  {https://doi.org/10.1103/PhysRevE.109.014125} {\bibfield  {journal} {\bibinfo
   {journal} {Phys. Rev. E}\ }\textbf {\bibinfo {volume} {109}},\ \bibinfo
  {pages} {014125} (\bibinfo {year} {2024}{\natexlab{b}})}\BibitemShut
  {NoStop}%
\bibitem [{\citenamefont {Goodfellow}\ \emph {et~al.}(2016)\citenamefont
  {Goodfellow}, \citenamefont {Bengio},\ and\ \citenamefont
  {Courville}}]{GoodBengCour16}%
  \BibitemOpen
  \bibfield  {author} {\bibinfo {author} {\bibfnamefont {I.~J.}\ \bibnamefont
  {Goodfellow}}, \bibinfo {author} {\bibfnamefont {Y.}~\bibnamefont {Bengio}},\
  and\ \bibinfo {author} {\bibfnamefont {A.}~\bibnamefont {Courville}},\
  }\href@noop {} {\emph {\bibinfo {title} {Deep Learning}}}\ (\bibinfo
  {publisher} {MIT Press},\ \bibinfo {address} {Cambridge, MA, USA},\ \bibinfo
  {year} {2016})\ \bibinfo {note}
  {\url{http://www.deeplearningbook.org}}\BibitemShut {NoStop}%
\bibitem [{\citenamefont {Dumoulin}\ and\ \citenamefont
  {Visin}(2018)}]{dumoulin2018guide}%
  \BibitemOpen
  \bibfield  {author} {\bibinfo {author} {\bibfnamefont {V.}~\bibnamefont
  {Dumoulin}}\ and\ \bibinfo {author} {\bibfnamefont {F.}~\bibnamefont
  {Visin}},\ }\href@noop {} {\bibinfo {title} {A guide to convolution
  arithmetic for deep learning}} (\bibinfo {year} {2018}),\ \Eprint
  {https://arxiv.org/abs/1603.07285} {arXiv:1603.07285 [stat.ML]} \BibitemShut
  {NoStop}%
\bibitem [{\citenamefont {Opitz}\ and\ \citenamefont
  {Maclin}(1999)}]{Opitz1999}%
  \BibitemOpen
  \bibfield  {author} {\bibinfo {author} {\bibfnamefont {D.}~\bibnamefont
  {Opitz}}\ and\ \bibinfo {author} {\bibfnamefont {R.}~\bibnamefont {Maclin}},\
  }\bibfield  {title} {\bibinfo {title} {Popular ensemble methods: An empirical
  study},\ }\href {https://doi.org/10.1613/jair.614} {\bibfield  {journal}
  {\bibinfo  {journal} {Journal of Artificial Intelligence Research}\ }\textbf
  {\bibinfo {volume} {11}},\ \bibinfo {pages} {169–198} (\bibinfo {year}
  {1999})}\BibitemShut {NoStop}%
\bibitem [{Note1()}]{Note1}%
  \BibitemOpen
  \bibinfo {note} {See Supplemental Material at [URL will be inserted by
  publisher] for details about the machine learning architecture, the
  renormalization group, and the data analysis}\BibitemShut {NoStop}%
\bibitem [{\citenamefont {Wang}\ and\ \citenamefont
  {Swendsen}(1988{\natexlab{b}})}]{PhysRevB.38.9086}%
  \BibitemOpen
  \bibfield  {author} {\bibinfo {author} {\bibfnamefont {J.-S.}\ \bibnamefont
  {Wang}}\ and\ \bibinfo {author} {\bibfnamefont {R.~H.}\ \bibnamefont
  {Swendsen}},\ }\bibfield  {title} {\bibinfo {title} {Monte carlo and
  high-temperature-expansion calculations of a spin-glass effective
  hamiltonian},\ }\href {https://doi.org/10.1103/PhysRevB.38.9086} {\bibfield
  {journal} {\bibinfo  {journal} {Phys. Rev. B}\ }\textbf {\bibinfo {volume}
  {38}},\ \bibinfo {pages} {9086} (\bibinfo {year}
  {1988}{\natexlab{b}})}\BibitemShut {NoStop}%
\end{thebibliography}%

\end{document}